# Current-Induced Spin Polarization in Gallium Nitride

W. F. Koehl, M. H. Wong, C. Poblenz, B. Swenson, U. K. Mishra, J. S. Speck, D. D. Awschalom[a]

*Center for Spintronics and Quantum Computation, University of California, Santa Barbara, California 93106*

**Electrically generated spin polarization is probed directly in bulk GaN using Kerr rotation spectroscopy. A series of n-type GaN epilayers are grown in the wurtzite phase both by molecular beam epitaxy (MBE) and metalorganic chemical vapor deposition (MOCVD) with a variety of doping densities chosen to broadly modulate the transverse spin lifetime, $T_2^*$. The spin polarization is characterized as a function of electrical excitation energy over a range of temperatures. Despite weak spin-orbit interactions in GaN, a current-induced spin polarization (CISP) is observed in the material at temperatures of up to 200 K.**

[a] Electronic mail: awsch@physics.ucsb.edu

Phenomena that can be used to initialize, control, or detect spins in condensed matter systems are of central importance to the field of spintronics. Of particular interest in semiconductors are mechanisms that allow these tasks to be completed solely by electrical means, since mature electronic technologies may potentially then be leveraged to implement spin-based devices.[1] Current-induced spin polarization (CISP) is one such phenomenon in which the charge carriers of a semiconductor become spin polarized as a current is made to flow through the material. In experiments characterizing CISP in bulk films, an electrical bias is applied across a semiconductor channel, which then causes a spin polarization to be generated along an axis that is in the plane of the film and perpendicular to the direction of current flow. The spin polarization generated is spatially uniform throughout the channel rather than accumulating only at the channel edges as it does in the spin Hall effect.

Although CISP has been observed in a number of two-[2,3,4] and three-[5,6] dimensional systems, its physical origins are not well understood. Spin-orbit coupling (SOC) is often used to explain the source of the spin polarization, but models that rely on SOC caused by electric fields within devices have not fully explained measurements in bulk epilayers. Others have suggested that charge carriers may become spin-polarized even in systems where SOC is absent,[7] but measurements of CISP so far have been in devices where spin-orbit interactions are relatively large. In an effort to more fully understand the role of spin-orbit interactions in CISP, we investigate the phenomenon in bulk n-type gallium nitride (GaN), a III-V system with significant technological interest for which SOC is expected to be small.[8]

The n-type GaN epilayers under study include a series of three grown via plasma-assisted molecular beam epitaxy (MBE) (samples A, B, and C) and a fourth grown via metalorganic chemical vapor deposition (MOCVD) (sample D). The MBE growths were deposited on semi-insulating 4H-SiC substrates under gallium rich conditions,[9] and a previously developed two-step buffer growth method was used[10] in which an initial 50 nm AlN nucleation layer was followed by a nominally-undoped 100 nm layer of GaN. After this, a final 515 nm n-type GaN layer was deposited which was doped with Si to achieve carrier concentrations of 4.5e17, 1.1e18, and 1.5e18 cm$^{-3}$ for samples A, B, and C respectively. The MOCVD growth began with a thin, low-temperature seed layer of GaN grown on sapphire,[11] followed by two compensation-doped GaN:Fe buffer layers of total thickness 850 nm.[12] A nominally-undoped 550 nm buffer layer of GaN was then deposited, after which a 1000 nm Si-doped active layer was grown that was measured to have a carrier concentration of 4.14e17 cm$^{-3}$. The dislocation densities of the MBE and MOCVD growths were measured via x-ray diffraction to be ~2e10 cm$^{-2}$ and ~1e9 cm$^{-2}$ respectively.[13]

Hall bar devices were fabricated along both the [11$\bar{2}$0] and [1$\bar{1}$00] directions of each sample with a reactive ion etch, after which ohmic contact pads with a Ti/Al/Ni/Au layer structure were deposited and subjected to a rapid thermal anneal at 870 °C for 30 seconds in N$_2$.[14] A Hall bar geometry, shown in Fig. 1(a), was chosen over a simpler design with only two contacts so that any device heating under an electrical bias could be quantified via measurements of the device resistance.

Devices were mounted in the variable temperature insert of a magneto-optical cryostat with the device channel parallel to the external magnetic field. An alternating square wave electrical bias with a frequency of 1307Hz was applied across the channel and varied between 0 and 20 Volts peak-to-peak, the latter of which corresponds to an electric field in the channel of roughly 66.7 mV/μm. We measured time-resolved Kerr rotation (TRKR) in the Voigt geometry to determine the electron spin coherence time ($T_2^*$) and g-factor ($g$) of each device as a function of temperature ($T$) and electric field ($E$).[15] The output from a mode-locked pulsed Ti:sapphire laser was frequency-doubled and split into a 1.5 mW circularly-polarized pump beam and a 0.15 mW linearly-polarized probe beam. These beams were focused onto the sample with a spot diameter of ~19 μm, and their wavelength ranged from 355.5 nm to 364.5 nm depending on $T$.

The g-factor was found to be insensitive to $T$, $E$, and doping density, and on average was measured to be $g$ = 1.96±0.04. In addition, at $T$ = 50 K, $B$ = 0.125 T, and $E$ = 0 mV/μm, $T_2^*$ was measured to be 978 ± 67, 238 ± 11, 169 ± 12, and 579 ± 21 ps for samples A,B,C, and D respectively. In sample A, $T_2^*$ at $T$ = 300 K was measured to be 104 ± 15 ps when $B$ = 0.5 T and $E$ = 0 mV/μm. This is a factor of three larger than the value of 35 ps reported in a previous study of bulk n-GaN,[16] and brings this figure of merit for GaN into the same order of magnitude as that seen in bulk n-type ZnO (190 ps)[17] and GaAs (110 ps).[18] This is despite having a dislocation density that is many orders of magnitude larger than that of state-of-the-art bulk GaAs.[19]

We then investigated whether CISP could be observed in any of the four GaN samples. We blocked the pump beam in our setup to avoid polarizing spins optically, increased the probe power to 1.5 mW, and measured the resultant KR signal via a single lockin referenced to the frequency of the electrical bias. The KR measured is proportional to the time-averaged projection of spin polarization along the direction of probe propagation. If spin polarization is generated along $y$ (Fig. 1(a)), the combined effects of spin precession and decoherence result in plots of KR vs. magnetic field having an odd-Lorentzian lineshape determined by the equation:

$$\theta_K = \theta_{el}\omega_L\tau/((\omega_L\tau)^2+1) \qquad (1)$$

where $\omega_L = g\mu_B B/\hbar$ is the Larmor precession frequency, $\mu_B$ is the Bohr magneton, $\hbar$ is the Plank constant, $\tau$ is the inhomogeneous transverse spin lifetime, and $\theta_{el}$ is the magnitude of the electrically generated KR angle.[5]

At $T = 50$ K, CISP was observed in samples A and D, but not in B or C. We then characterized CISP as a function of $E$ and $T$ in samples A and D. A typical measurement of CISP in GaN can be seen in Fig. 1(b). For each value of $E$, a scan such as this was taken and then fit to Eq. (1) using the value of $g$ found from TRKR measurements taken under the same experimental conditions. From this, values for $|\theta_{el}|$ and $\tau$ were extracted. The polarity of $\theta_{el}$ could not be determined because attempts to measure nuclear polarization in the epilayers were unsuccessful. Plots of $|\theta_{el}|$ and $\tau$ as a function of $E$ across the [1$\bar{1}$00] devices of samples A and D are shown in Fig. 2 for several different temperatures. Large error bars are due to a small signal-to-noise ratio, as the change in KR measured at low biases and high temperatures was often close to the resolution of our setup (~80nrad). The signal was in general stronger in sample D, and data from sample A where $E < 20$ mV/μm and $T > 50$ K could not be fit because the signal dropped below the noise floor. In both samples at all temperatures, $|\theta_{el}|$ increases with $E$, indicating that more charge carriers become spin polarized as the bias is raised.

At 50 K, $\tau$ decreases dramatically as a function of $E$ in both samples. As $T$ is raised this behavior becomes progressively less pronounced so that at the highest temperatures $\tau$ does not vary with $E$ at all. In Fig. 3(a-b) we plot $T_2^*$ as a function of $E$ and $T$ as measured via TRKR and compare with the appropriate plots of $\tau$ in Fig. 2. $T_2^*$ exhibits the same behavior as $\tau$ in both samples at 50 K, while at higher temperatures it remains essentially constant in $E$. In sample D, the similarity between $T_2^*$ and $\tau$ is particularly striking, as they remain close in magnitude throughout the data set. The electronic mobility of the material is low (~100 cm$^2$/V·s for sample A at 50 K), and spatially-resolved TRKR measurements revealed no distortion of optically-induced spin packets under electrical biases. Therefore, the behavior of $T_2^*$ at 50 K is not due to spin transport effects, which suggests that device heating is largely responsible for the decrease in both $T_2^*$ and $\tau$ as a function of $E$ at low temperatures. Measurements of device resistance taken at each value of $T$ and $E$ corroborate this picture. Using device resistance as a proxy for device temperature, the heating in sample A at 50 K and $E$ = 66.7 mV/μm was found to be ~35 K. This is reasonable considering that devices on sample A are ~20 kΩ at 50 K and zero bias. In Fig. 3(c) we plot $T_2^*$ as a function of $T$ at zero bias for samples A and D and see that the projected value of $T_2^*$ at 85 K is in rough agreement with the value measured at $T$ = 50 K and $E$ = 66.7 mV/μm. From further analysis of transport and $T_2^*$ data we find that heating in sample A is less than 9 K at 100 K and less than 6 K above 150 K, and that heating is similar in sample D.

The exact relationship between SOC and CISP is unclear. Previous attempts to associate CISP in bulk epilayers with a Rashba-like strain-induced spin splitting have been unsuccessful.[5] In (110) AlGaAs quantum wells, CISP was argued to be the result of Dresselhaus SOC exclusively,[2] while in other two-dimensional systems, Rashba SOC was invoked to explain the effect.[3,4] Most recently, CISP was observed in ZnSe at temperatures of up to 295 K even though no SO effective magnetic fields were measured within the material.[6] GaN is a material for which SOC coefficients are predicted to be small. In particular, the k-cubic Dresselhaus coefficient has been calculated via a Kane-type **k·p** model to be 0.32 eV·Å$^3$,[8] while measurements in AlGaN/GaN two-dimensional electron gases indicate that the upper bound for the Rashba coefficient is in the range of ~1-3 e·Å$^2$.[20,21] Despite small SOC coefficients, CISP can be clearly resolved in the material at temperatures of up to 200 K. Furthermore, the data in Fig. 2 behave similarly to that of previous studies of [1$\bar{1}$0] InGaAs channels,[5] suggesting that the mechanism behind CISP in GaN is similar in nature to that of previously explored bulk materials.

No changes in $|\theta_{el}|$ or $\tau$ were observed when CISP was characterized in devices oriented along the [11$\bar{2}$0] direction. This is in contrast to previous results in InGaAs, for which $|\theta_{el}|$ varied strongly in devices oriented 90° relative to one-another. It is possible that this discrepancy could be due to the difference in symmetries of Dresselhaus SOC in wurtzite and zinc blende materials. The relative polarity of zinc blende Dresselhaus and Rashba fields for electron spins travelling along the [110] direction is opposite that of spins moving along the [1$\bar{1}$0] axis.[22] However, in wurtzite materials, Dresselhaus and Rashba terms possess the same symmetry for electron spins travelling in any direction parallel to the c-plane.[8] A more detailed study of the crystallographic dependence of CISP in wurtzite GaN is needed to determine whether this is in fact what is occurring.

Above 200 K, both $\theta_{el}$ and $\tau$ become very small, so that the odd-Lorentzian lineshape characteristic of CISP becomes broad and flat and therefore experimentally difficult to resolve. However, the high dislocation densities of GaN limit the spin coherence times of the material,[23] and high-purity GaN is predicted to possess electron spin coherence times about two orders of magnitude larger than those of GaAs at room temperature, in part due to the weak SOC of the material.[24] Therefore, with further development of growth techniques, it is possible that the observation of CISP in GaN may be extended to even higher temperatures.

We thank ARO and ONR for financial support. Devices were fabricated in the UCSB nanofabrication facility, part of the NSF funded NNIN network.

[24] Z. G. Yu, S. Krishnamurthy, M. V. Schilfgaarde, and N. Newman, Phys. Rev. B **71**, 245312 (2005).
11

**Figure Captions**

**FIG. 1.** **(a)** Device schematic. Contacts are yellow and the channel and sidearms are grey. ***B*** and ***E*** are parallel to the channel for all measurements. **(b)** $\theta_K$ as a function of $B$ for n-GaN at 50 K with $E = 66.7$ mV μm$^{-1}$. Channel is oriented along the [1$\bar{1}$00] axis of sample D. Black circles are data while the blue line is a fit to Eq. (1).

**FIG. 2.** Electric field dependence of $|\theta_{el}|$ and $\tau$ at various temperatures in devices oriented along [1$\bar{1}$00] in samples A and D. **(a)** $|\theta_{el}|$ and **(b)** $\tau$ as extracted from fits to data taken in sample D at 50 K (black squares), 100 K (red circles), and 150 K (blue upturned triangles). **(c)** $|\theta_{el}|$ and **(d)** $\tau$ from sample A at 50 K (black squares). **(e)** $|\theta_{el}|$ and **(f)** $\tau$ from sample A at 100 K (red circles), 150 K (blue upturned triangles), and 200 K (green downturned triangles). 50 K data from sample A plotted separately because larger scale is needed for $\tau$.

**FIG. 3.** Electric field dependence of $T_2^*$ as measured in devices oriented along [1$\bar{1}$00] in samples **(a)** A and **(b)** D at 50 K (black squares), 100 K (red circles), 150 K (blue upturned triangles), and 200 K (green downturned triangles). **(c)** $T_2^*$ as a function of temperature in samples A (black symbols) and D (red symbols). From fits to TRKR data taken at $B = 0.125$ T (squares), 0.25 T (circles), 0.5 T (upturned triangles), and 1.0 T

(downturned triangles). **(d)** $\theta_K$ as a function of $B$ for n-GaN at 200 K. Channel is oriented along the [1$\bar{1}$00] axis of sample A and $E = 66.7$ mV µm$^{-1}$. Black circles are data while the blue line is a fit to Eq. (1). $\theta_K$ measurement error is roughly equal to point scatter in data.

(a) 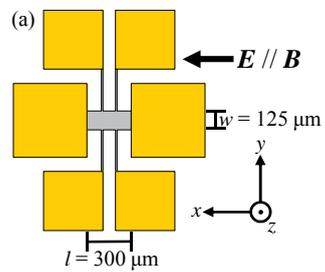
(b) 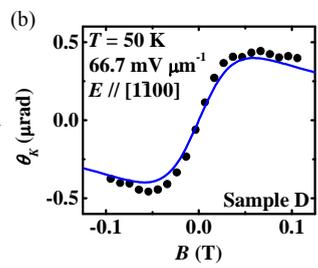

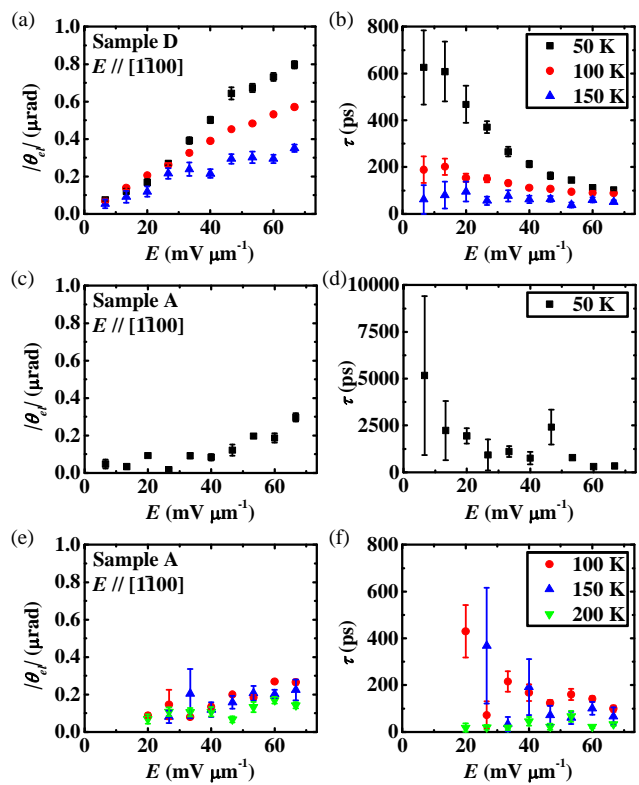

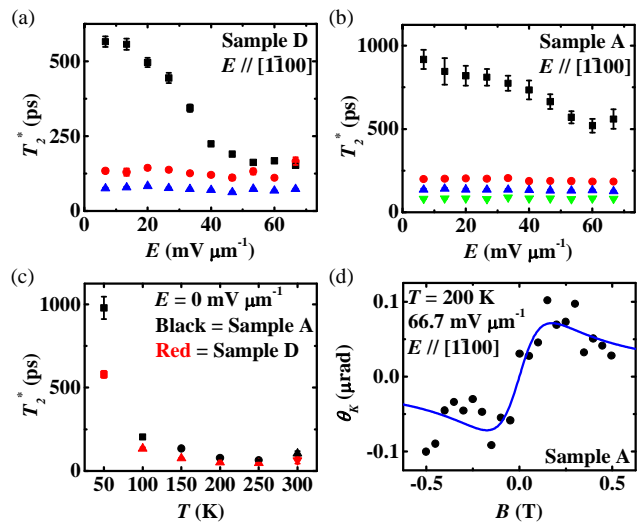